\documentclass{PoS}

\usepackage{amsmath}
\usepackage{upgreek}

\newcommand\Bezier{B\'{e}zier}
\newcommand\ROBAST{\texttt{ROBAST}}

\title{Prototyping of Hexagonal Light Concentrators for the Large-Sized Telescopes of the Cherenkov Telescope Array}

\ShortTitle{LST Light Concentrators for CTA}

\author{\speaker{Akira~Okumura}$^{a,b}$, Sakiya~Ono$^c$, Syunya~Tanaka$^c$, Masaaki~Hayashida$^d$, Hideaki~Katagiri$^c$, and Tatsuo~Yoshida$^c$, for the CTA Consortium\footnote{Full consortium author list at http://cta-observatory.org/}\\
  \llap{$^a$}Solar-Terrestrial Environment Laboratory, Nagoya University, Furo-cho, Chikusa-ku, Nagoya, Aichi 464-8601, Japan\\
  \llap{$^b$}Max-Planck-Institut f\"{u}r Kernphysik, P.O. Box 103980, D 69029 Heidelberg, Germany\\
  \llap{$^c$}College of Science, Ibaraki University, 2-1-1, Bunkyo, Mito 310-8512, Japan\\
  \llap{$^d$}Institute for Cosmic Ray Researcht, University of Tokyo, 5-1-5 Kashiwanoha, Kashiwa, Chiba 277-8582, Japan\\
  E-mail: \email{oxon@mac.com} (A.~O.)}

\abstract{Reflective light concentrators with hexagonal entrance and exit apertures are frequently used at the focal plane of gamma-ray telescopes in order to reduce the size of the dead area caused by the geometries of the photodetectors, as well as to reduce the amount of stray light entering at large field angles. The focal plane of the large-sized telescopes (LSTs) of the Cherenkov Telescope Array (CTA) will also be covered by hexagonal light concentrators with an entrance diameter of 50 mm (side to side) to maximize the active area and the photon collection efficiency, enabling realization of a very low energy threshold of 20 GeV. We have developed a prototype of this LST light concentrator with an injection-molded plastic cone and a specular multilayer film. The shape of the plastic cone has been optimized with a cubic \Bezier\ curve and a ray-tracing simulation. We have also developed a multilayer film with very high reflectance ($\gtrsim95$\%) along wide wavelength and angle coverage. The current status of the prototyping of these light concentrators is reported here.}

\FullConference{The 34th International Cosmic Ray Conference,\\
		30 July- 6 August, 2015\\
		The Hague, The Netherlands}

\begin{document}

\section{Introduction}
The focal planes of imaging atmospheric Cherenkov telescopes (IACTs) are usually covered with light concentrators with hexagonal entrance and exit apertures and specular inner surfaces. Light concentrators reduce the size of the dead area between the photodetectors and reduce the number of direct and indirect night sky background photons entering at large field angles, increasing the signal-to-noise ratio for faint Cherenkov photons. Compound parabolic cones (CPCs, also referred to as ``Winston cones'') \cite{Winston:1970:Light-Collection-within-the-Fr} and similar designs have been commonly used for this purpose \cite{Kabuki:2003:Development-of-an-atmospheric-Cherenkov-,Bernlohr:2003:The-optical-system-of-the-H.E.S.S.-imagi,Nagai:2008:Focal-Plane-Instrumentation-of-VERITAS}, because an ideal CPC for a two-dimensional space has a collection efficiency of 100\% and 0\% for photons entering at field angles less than the cutoff angle ($\theta < \theta_\mathrm{cutoff}$) and for those entering at field angles greater than the cutoff angle ($\theta > \theta_\mathrm{cutoff}$), respectively. The inner curved surfaces of such a light concentrator are usually coated with aluminum and protection layers, ultimately achieving specular reflection, with a reflectance of ${\sim}90$\% in the wavelength band $300$--$600$~nm.

In the past decade, one of big challenges in IACT development was to achieve low energy thresholds below $100$~GeV, and thus telescopes with large-diameter mirrors have been built for the MAGIC ($\phi17$~m, $>25$~GeV) and H.E.S.S. II ($\phi28$~m, $>30$~GeV) telescopes. To lower the energy threshold further with a reasonable construction cost for IACT optical systems, it is necessary to develop a light concentrator with a higher collection efficiency.

The large-sized telescopes (LSTs, $\phi23$~m) in the Cherenkov Telescope Array (CTA), the next-generation very-high-energy gamma-ray observatory, are currently under development, and will allow observation in a low-energy band ($20$--$200$~GeV) of the gamma-ray sky in both the northern and southern hemispheres \cite{Actis:2011:Design-concepts-for-the-Cherenkov-Telesc,Acharya:2013:Introducing-the-CTA-concept}. Their light concentrators play a crucial role in realizing this challenging energy threshold by increasing the effective photon collection area of the LSTs. Some new techniques we have developed for a prototype of the LST light concentrators are reported in the present paper.

\section{Prototyping of LST Light Concentrators}
\subsection{The LST Optical System and Camera}

The LST optical system comprises a parabolic mirror and a flat focal plane. The mirror is composed of 198 segmented spherical mirror facets to approximate a large parabolic mirror with a focal length $f$ of $28$~m and a diameter $D$ of $23$~m. The LST camera, which is an array of 1855 photomultiplier tubes (PMTs), is placed at the focal plane, and each PMT pixel is equipped with a hexagonal light concentrator with an entrance diameter of $50$~mm (side to side). The light concentrators need to efficiently collect photons with field angles up to ${\sim}25^\circ$. This is because the viewing angle from the camera center is $\tan^{-1}(D/2f)=22.3^\circ$.

\subsection{The Prototype Design}

Our proposed LST light concentrator prototype design is similar to that of a hexagonal CPC. However, a few new ideas have been employed to increase the collection efficiency and achieve the LST threshold.

\begin{figure}
  \centering
  \includegraphics[width=\columnwidth]{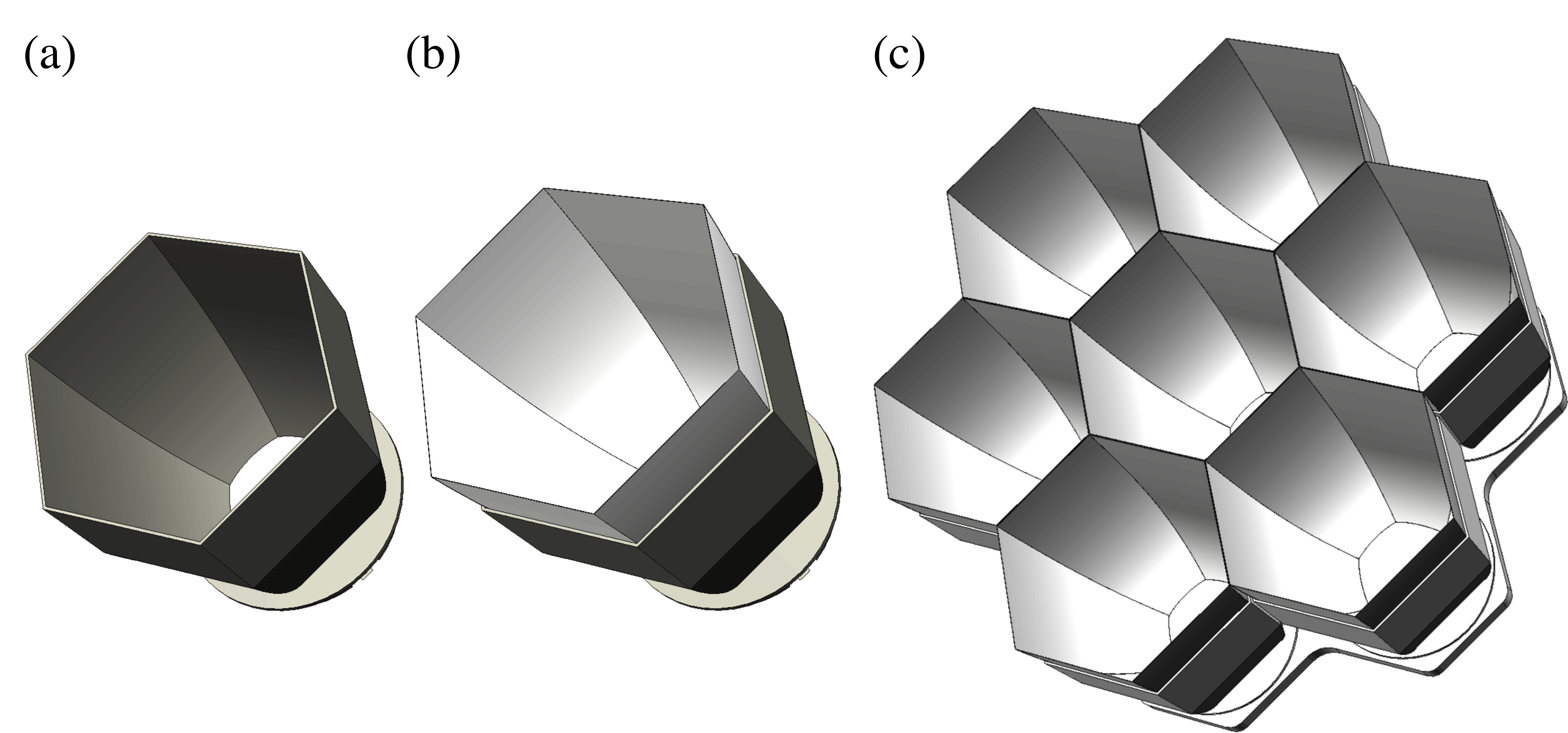}
  \caption{(a) 3D CAD model of the base ABS cone of an LST light concentrator. (b) Same as (a), but six specular films are attached to the cone. (c) A light concentrator cluster comprised of seven copies of (b) and an interface plate at the bottom.}
  \label{fig:CAD}
\end{figure}

We use multilayer optical films for the inner surfaces of the light concentrator. As illustrated in Figure~\ref{fig:CAD}, six film pieces are glued onto an acrylonitrile butadiene styrene (ABS) plastic cone made by injection molding, but the upper end is kept free (i.e., the plastic cone height is shorter than the film length). This design has three advantages: Very high reflectance, a smaller dead area, and that it is non-conducting.

The film has higher reflectance ($95$--$100$\%) than aluminum or any other metal coating ($<95$\%) in a broad range of wavelengths ($300$--$700$~nm) and angles of incidence ($20^\circ$--$70^\circ$) simultaneously, and thus the collection efficiency is expected to be improved by at least a few percent. We can also reduce the dead area between adjacent pixels as the film thickness is only about $80$~$\upmu$m. In contrast, a plastic cone with aluminum coating has an inevitable dead area obtained from the minimum thickness of the injection-molded plastic (typically ${\sim}0.5$~mm). Thus, in the case of a $50$~mm entrance diameter, the ratio of the effective entrance area to the pixel area becomes $99.2$\% ($=(49.8/50.0)^2$) and $96.0$\% ($=(49.0/50.0)^2$) for the multilayer film and aluminum coating solutions, respectively. The total collection efficiency we will obtain using multilayer films is expected to be $5$\%--$10$\% higher than that of an aluminum-coated light concentrator.

We use a cubic \Bezier\ curve to form the six curved inner walls. This results in a sharp cutoff with a lower inner surface area (i.e., a shorter cone height than for a normal CPC), and so we can reduce the production cost of the multilayer films, which would be the dominant cost in mass production. We have extended the original idea of using a \Bezier\ curve for light concentrators \cite{Okumura:2012:Optimization-of-the-collection-efficiency-of-a-hex} by introducing two additional parameters: the cone height and exit aperture diameter\footnote{The entrance diameter was fixed to $49.8$~mm assuming a film thickness of $100$~$\upmu$m.}.

We simulated the collection efficiency of various light concentrator designs with different parameters: the coordinates of the \Bezier\ curve control points (see Figure~2(c) in \cite{Okumura:2012:Optimization-of-the-collection-efficiency-of-a-hex}), the exit aperture diameter, and the cone height. The simulation was performed using \ROBAST, an open-source non-sequential ray-tracing simulation library \cite{Okumura:2011:Development-of-Non-sequential-Ray-tracing-Software,Okumura:2015:ROBAST:-Development-of-a-ROOT-Based-Ray-Tracing-Li}, assuming that the inner surfaces have a constant reflectance of $95$\% that is independent of the wavelength or angle of incidence. The cone height and diameter of the exit aperture of the prototype shape are $68.0$ and $11.0$~mm, respectively.

\begin{figure}
  \begin{minipage}[t]{0.5\hsize}
    (a)\\
    \includegraphics[width=\hsize]{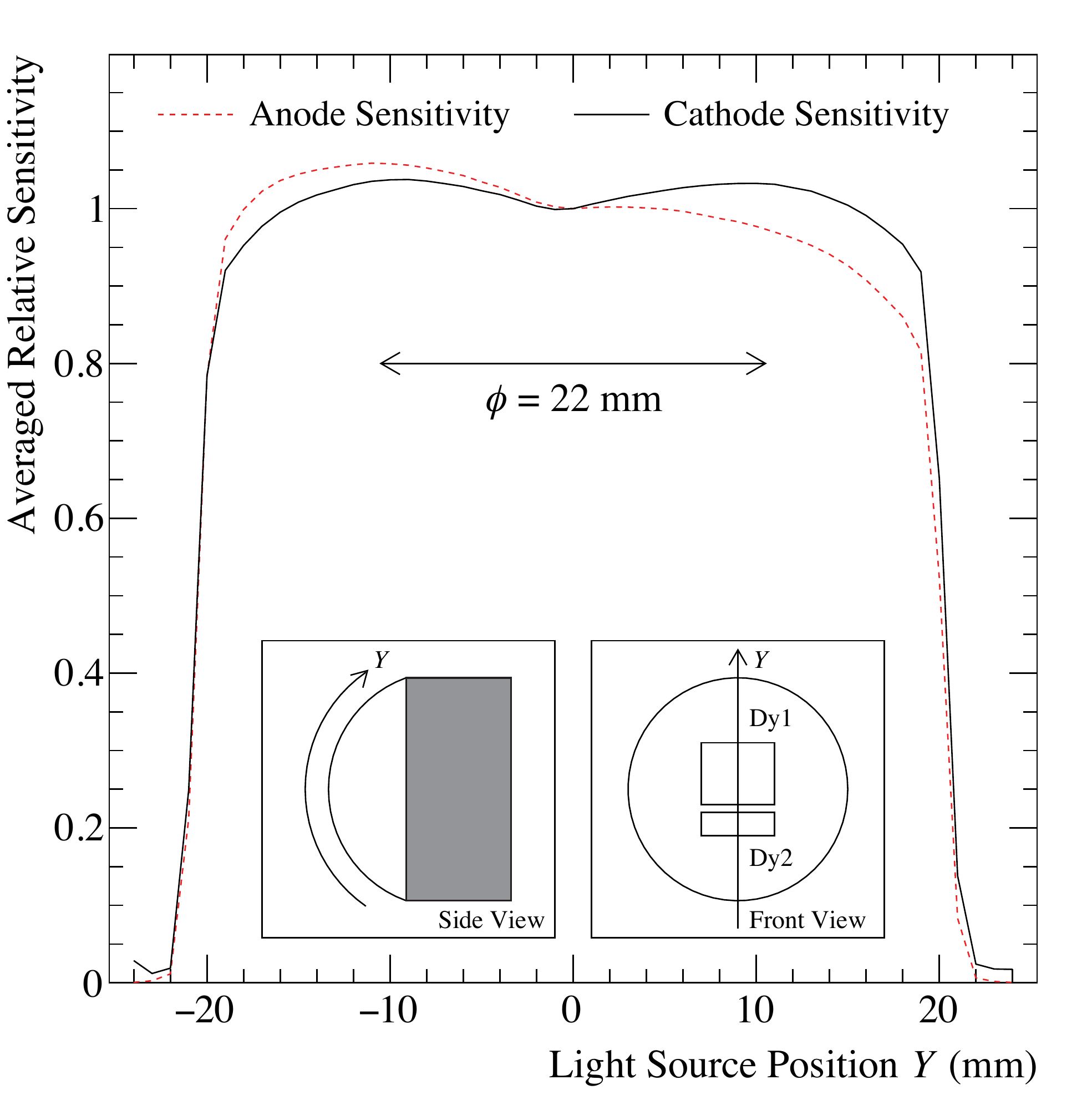}
  \end{minipage}
  \begin{minipage}[t]{0.5\hsize}
    (b)\\
    \includegraphics[width=\hsize]{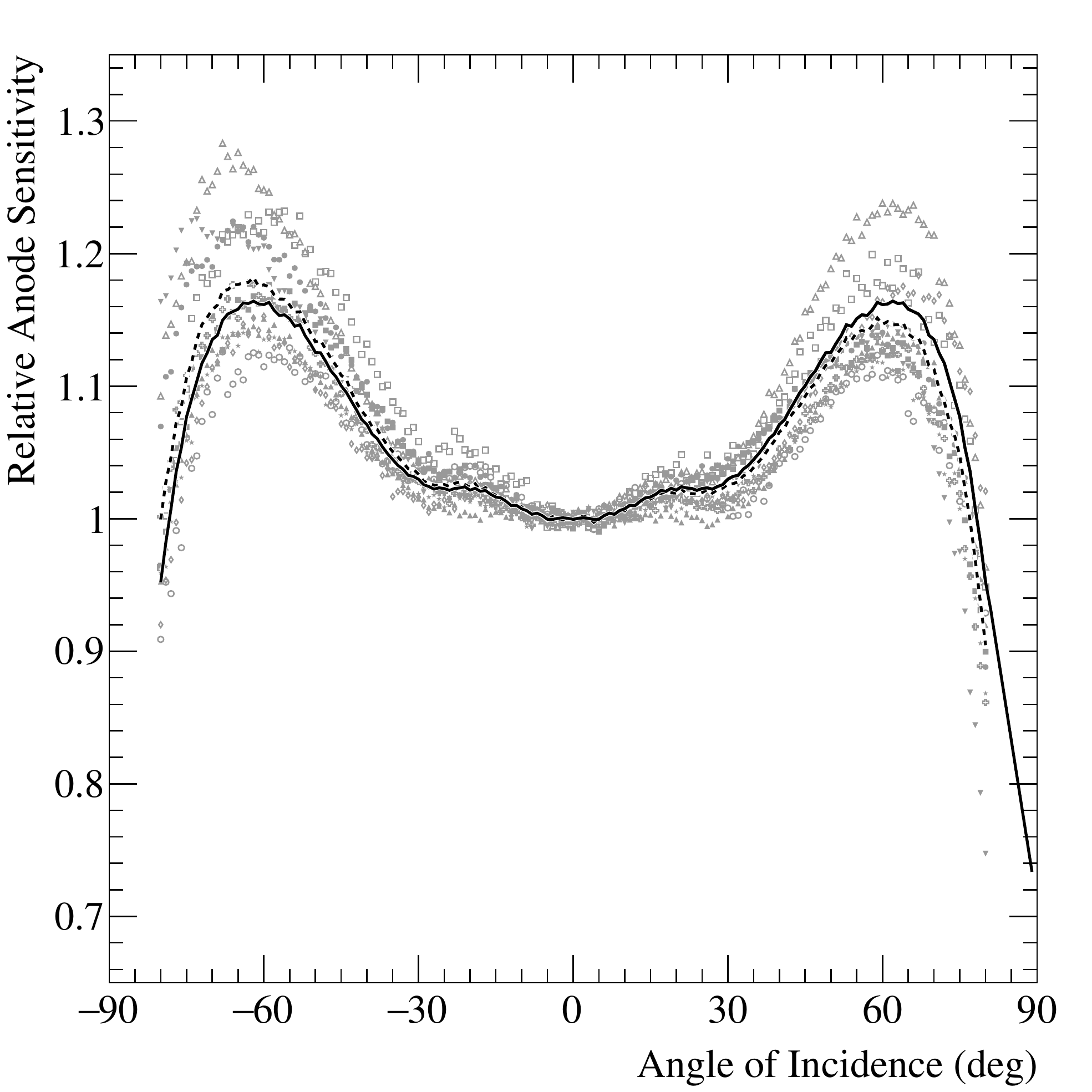}
  \end{minipage}
  \caption{(a) The averaged anode (red dashed) and cathode (black solid) sensitivity of the LST PMTs (Hamamatsu Photonics R11920-100) versus light source position. Measurements of 90 PMTs were averaged and normalized to the position at $Y=0$~mm. The insets schematically illustrate a PMT and the definition of the curved $Y$ axis. (b) The photon detection efficiency versus angle of incidence, normalized relative to the vertical (i.e., $0$~deg). The data points show measured values for eight PMTs. The dashed line shows the average of the eight PMTs. The solid line shows the symmetrical average (and extrapolates to $80$--$90^\circ$).}
  \label{fig:sensitivity}
\end{figure}

In the \ROBAST\ ray-tracing simulation, we have also taken into account the position and angular dependence of the PMT sensitivity to reproduce the response of a light concentrator (the anode sensitivity versus field angle) as accurately as possible. This is because it has been reported that a PMT with a hemispherical and matte input window has a strong position and angular dependence in its anode sensitivity \cite{Paneque:2003:A-method-to-enhance-the-sensitivity-of-photomultip}, and therefore a significant effect appears in the light concentrator response. Figure~\ref{fig:sensitivity} shows the position and angular dependence of the anode sensitivity of LST PMTs.

\subsection{Development of a Specular Multilayer Film}

We have developed a specular multilayer film to achieve very high reflectance over a wide range of wavelengths and angles of incidence simultaneously. Cherenkov photons arriving at the focal plane have wavelengths of ${\sim}300$~nm or longer, and they are reflected on the inner surfaces of a light concentrator at angles of incidence ranging from $35^\circ$ to $90^\circ$. To realize very high reflectance in this wavelength range and very high angle coverage, we have developed a new film by coating an existing specular film product\footnote{Multilayer polyester film. Nominal reflectance in $400$--$800$ nm is $98$\% or more.} (Vikuiti ESR by 3M, hereafter ESR) with additional layers\footnote{By Bte Bedampfungstechnik GmbH in Elsoff, Germany.}.

\begin{figure}
  \centering
  \includegraphics[width=\columnwidth]{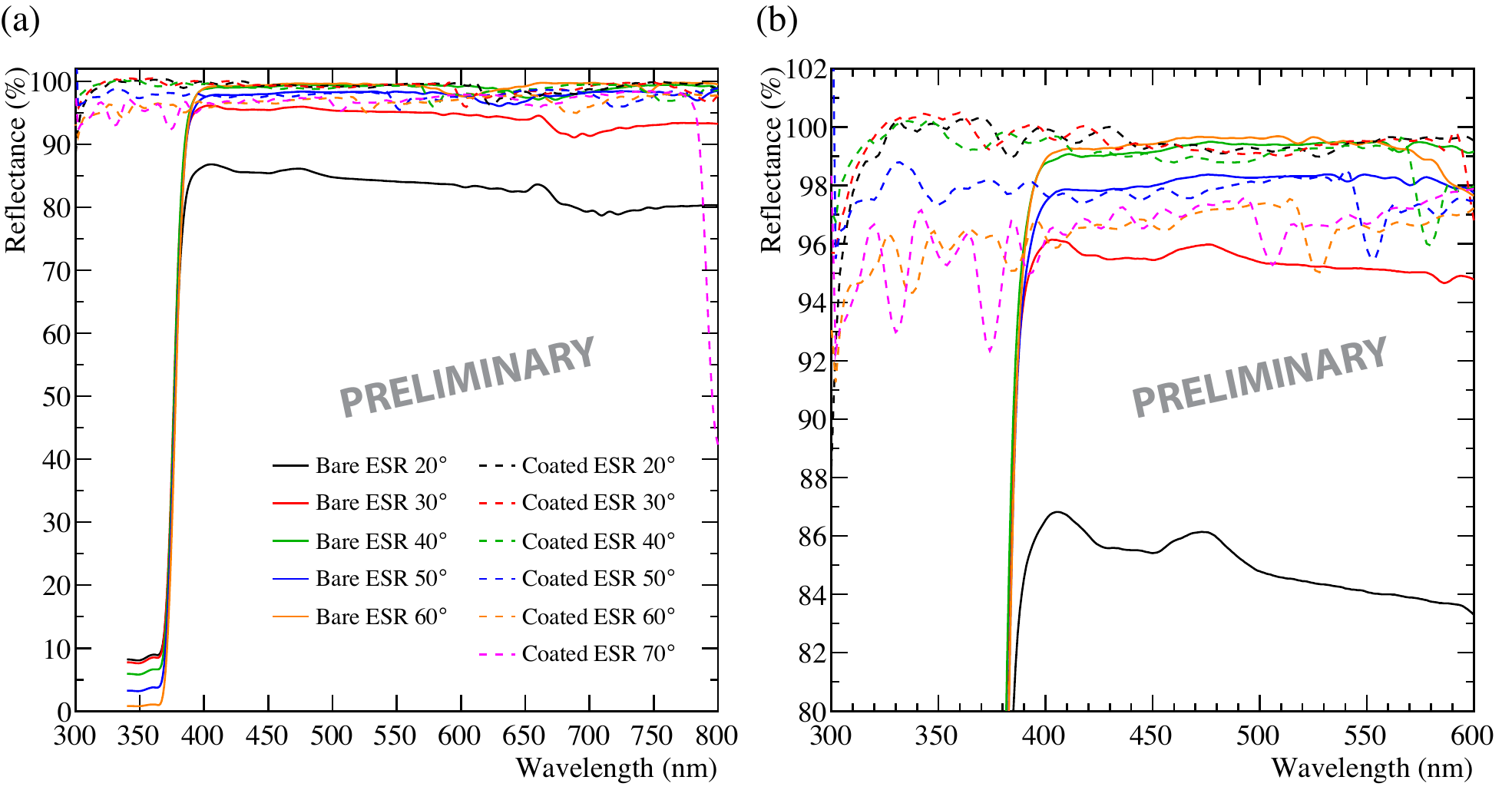}
  \caption{(a) The measured reflectance of bare (solid lines) and coated (dashed) ESR films. (b) Same as (a) but only limited axis ranges are shown. The systematic uncertainty of these measurements is estimated to be ${\sim}1$\%, and thus some data points are higher than $100$\%. }
  \label{fig:reflectance}
\end{figure}

ESR does not have high reflectance below $400$~nm due to the absorption of ultraviolet photons in the material. However, our additional layers greatly enhance the reflectance in the $300$--$400$~nm range, and the angular dependence of the bare ESR reflectance has been mitigated. Figure~\ref{fig:reflectance} compares the reflectance of bare and coated ESR films measured with various angles of incidence up to $70^\circ$.

\subsection{Preliminary Measurement Result}

We assembled several light concentrators with ABS plastic cones and coated ESR films as depicted in Figure~\ref{fig:CAD}. We have measured the relative anode sensitivity of a prototype for field angles from $-45^\circ$ to $+45^\circ$ with $1^\circ$ ($|\theta|\leq25^\circ$) and $0.5^\circ$ ($|\theta|\geq25^\circ$) steps. A PMT and attached light concentrator on a motorized rotation stage were illuminated with a blue light emitting diode. We here define the relative anode sensitivity $Q$ as
\begin{equation}
  Q = \frac{A_\mathrm{LC}(\theta)}{A_\mathrm{25\ mm}} \times \frac{1}{3.7515\times\cos\theta},
\end{equation}
where $A_\mathrm{LC}(\theta)$ is the PMT pulse area measured with a light concentrator and an oscilloscope at an field angle of $\theta$, $A_\mathrm{25\ mm}$ the pulse area measured with a hexagonal mask ($25$~mm diameter) at $\theta=0^\circ$, $\theta$ the field angle, and $3.7515$ the ratio of the entrance aperture area (a hexagon with a diameter of $50$~mm) to the $25$-mm mask\footnote{The mask area is slightly larger than a $25$-mm hexagon because of the hemispherical shape of the PMT, and thus $3.7515$ is used here instead of $4$.}. This value can be regarded as the collection efficiency of the light concentrator if the position and angular dependence of the PMT anode sensitivity (Figure~\ref{fig:sensitivity}) is negligible.

\begin{figure}
  \centering
  \includegraphics[width=0.5\columnwidth]{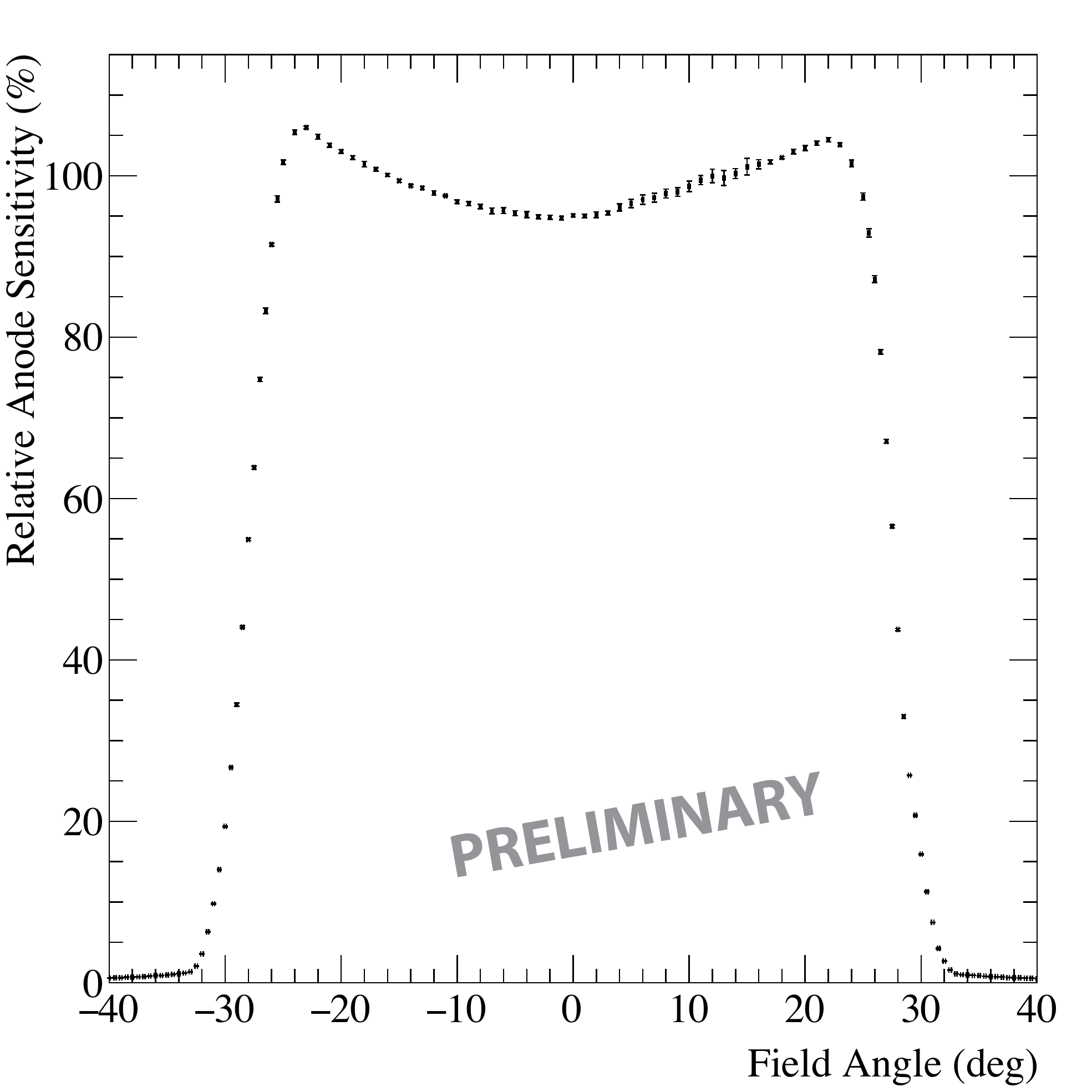}
  \caption{The measured relative anode sensitivity of a prototype LST light concentrator as a function of the field angle.}
  \label{fig:response}
\end{figure}

A preliminary measurement result of the prototype is shown in Figure~\ref{fig:response}. The measured response has a bump larger than $100$\% around $23^\circ$, which can be explained by large incidence angles on the PMT surface and the angular dependence of the PMT anode sensitivity shown in Figure~\ref{fig:sensitivity}(b). The relative anode sensitivity in the region of interest ($|\theta|<25^\circ$) is higher than $95$\%.

\section{Conclusion}

We have developed a prototype of an LST light concentrator using a specular multilayer film with very high reflectance in wide wavelength and angle ranges. The measured film reflectance is higher than that of normal aluminum coating, and thus the collection efficiency and the relative anode sensitivity of our design are higher than those of other light concentrators used in existing IACTs. Further measurements and the systematic study of our light concentrators will be reported in another upcoming paper.

\section*{Acknowledgments}
This study was supported by JSPS KAKENHI Grant Number 25707017. A.~O. was supported by a Grant-in-Aid for JSPS Fellows during his two-year visit to the University of Leicester, UK. The original data for Figure~\ref{fig:sensitivity}(a) was provided by Hamamatsu Photonics. We gratefully acknowledge support from the agencies and organizations listed under ``Funding Agencies'' at this website: http://www.cta-observatory.org/.

\bibliographystyle{JHEP}
\bibliography{oxon}


\end{document}